\documentstyle[aps,pra,eqsecnum,psfig,multicol,graphicx]{revtex}

\begin{document}
\draft
\title{Bayesian feedback versus Markovian feedback
in a two-level atom}
\author{H.~M. Wiseman${}^{1}$, 
Stefano~Mancini${}^{2}$, Jin Wang${}^{3}$}
\address{
${}^{1}$School of Science, Griffith University, Brisbane, Queensland
4111, Australia \\
${}^{2}$INFM, Dipartimento di Fisica, Universit\`a di Camerino,
I-62032 Camerino, Italy\\
${}^{3}$Department of Physics,
The University of Queensland, Brisbane, Queensland 4072, Australia
}
\date{\today}
\maketitle

\begin{abstract}
We compare two different approaches to the control of the dynamics of 
a continuously monitored open quantum system. The first 
is Markovian feedback as introduced in quantum optics by Wiseman and Milburn 
[Phys. Rev. Lett. {\bf 70}, 548 (1993)]. The second is feedback based 
on an estimate of the system state, developed recently by Doherty {\em 
et al.} [Phys. Rev. A {\bf 62}, 012105 (2000)]. Here we choose to call 
it, for brevity, {\em Bayesian feedback}. For systems with nonlinear 
dynamics, we 
expect these two methods of feedback control to give markedly 
different results. 
The simplest possible nonlinear system is a driven and 
damped two-level atom, so we 
choose this as our model system. 
The monitoring is taken to be homodyne detection of the atomic 
fluorescence, and the control is by modulating the driving. 
The aim of the feedback in both 
cases is to stabilize the internal state of the atom as close as 
possible to an 
arbitrarily chosen pure state, in the presence of inefficient 
detection and other forms of decoherence. 
Our results (obtain without recourse to stochastic simulations) prove 
that  Bayesian 
feedback is never inferior, and is usually superior, 
to Markovian feedback. However it would be far  more difficult to implement 
than Markovian feedback and it loses its superiority when obvious 
simplifying approximations are made. It is thus not clear which form 
of feedback would be better
 in the face of inevitable experimental imperfections.
\end{abstract}

\pacs{42.50.Lc, 42.50.Ct, 03.65.Ta}

\newcommand{\beq}{\begin{equation}}
\newcommand{\eeq}{\end{equation}}
\newcommand{\bqa}{\begin{eqnarray}}
\newcommand{\eqa}{\end{eqnarray}}
\newcommand{\nn}{\nonumber}
\newcommand{\nl}[1]{\nn \\ && {#1}\,}
\newcommand{\erf}[1]{Eq.~(\ref{#1})}
\newcommand{\erfs}[2]{Eqs.~(\ref{#1}) and (\ref{#2})}
\newcommand{\erft}[2]{Eqs.~(\ref{#1}) -- (\ref{#2})}
\newcommand{\rf}[1]{(\ref{#1})}
\newcommand{\dg}{^\dagger}
\newcommand{\rt}[1]{\sqrt{#1}\,}
\newcommand{\smallfrac}[2]{\mbox{$\frac{#1}{#2}$}}
\newcommand{\half}{\smallfrac{1}{2}}
\newcommand{\bra}[1]{\langle{#1}|}
\newcommand{\ket}[1]{|{#1}\rangle}
\newcommand{\ip}[2]{\langle{#1}|{#2}\rangle}
\newcommand{\schx}{Schr\"odinger }
\newcommand{\sch}{Schr\"odinger}
\newcommand{\heix}{Heisenberg }
\newcommand{\hei}{Heisenberg}
\newcommand{\bl}{{\bigl(}}
\newcommand{\br}{{\bigr)}}
\newcommand{\ito}{It\^o }
\newcommand{\str}{Stratonovich }
\newcommand{\dbd}[1]{\frac{\partial}{\partial {#1}}}
\newcommand{\sq}[1]{\left[ {#1} \right]}
\newcommand{\cu}[1]{\left\{ {#1} \right\}}
\newcommand{\ro}[1]{\left( {#1} \right)}
\newcommand{\an}[1]{\left\langle{#1}\right\rangle}
\newcommand{\implies}{\Longrightarrow}
\newcommand{\bfi}{{\bf I}_{[0,t)}}

\begin{multicols}{2}

\section{Introduction}

Quantum feedback arises when the environment of an open quantum
system is deliberately engineered so that information lost from the
system into that environment comes back to affect the system again.
Typically the environment is large and  would 
be regarded at least in part as a classical system. 
In the case where the system dynamics
are Markovian in the absence of feedback, the information lost to
the environment can be treated as classical information: the
measurement result. The feedback loop thus consists of a quantum
system, a classical detector (which turns quantum information into
classical information) and a classical actuator (which uses the
classical information to affect the quantum system).

In general quantum feedback is difficult to treat because any time
delay or filtering in the feedback loop makes the system
dynamics non-Markovian. A great simplification arises for Markovian
feedback, where the measurement results are used immediately to alter
the system state, and may then be forgotten. In this case the dynamics
including
feedback may be described by a master equation in the Lindblad form.
This was shown by Wiseman and
Milburn \cite{WisMil93b,WisMil94a} for homodyne detection and Wiseman
\cite{Wis94a} in general. This description of feedback has been applied
to a wide variety of systems and for a wide variety of purposes (see
for example Refs~\cite{Wis95a,TomVit95,MabZol96,%
HorKil97,ManTom97,ManVitTom98,WisTho01}).

In a previous work \cite{WanWis01},
two of us applied the Wiseman-Milburn feedback
theory to show that almost \cite{fn1} all pure states of a fluorescent
two-level atom can be stabilized by Markovian feedback based on homodyne
detection of the fluorescence. That is, by adding an amplitude
 modulation
to the laser driving the atom proportional to the just-measured homodyne
photocurrent, the atom would obey a master equation having any given
pure state on the Bloch sphere as its stationary state. Without
feedback, the only pure stationary state is the ground state, in the
absence of driving. That work generalized the earlier results by
Hofmann, Mahler and Hess \cite{HofHesMah98,HofMahHes98} on the same
problem in a number of ways. One generalization was
to study the effect of a non-unit
efficiency of the homodyne detection. This was shown to be deleterious to the
maximum purity of the stationary states, especially those in the upper
half of the Bloch sphere.

For non-Markovian feedback, the master equation approach of Wiseman
and Milburn cannot be used. However, the formalism first used to derive
the Wiseman-Milburn master equation, quantum trajectories, {\em can} be used.
Quantum trajectories \cite{Car93} describe the stochastic evolution of
the state of an open quantum system conditioned upon the results of
measurements
performed upon its environment. 
They were first derived from abstract quantum measurement theory 
\cite{Bel88,BelSta92,Bar90,Bar93} but were independently invented in 
quantum optics for practical purposes \cite{DalCasMol92,GarParZol92,Car93}.
 In the special case where the system
has linear dynamics, the measurement is linear (e.g. homodyne
detection), and the feedback dynamics is linear, the quantum
trajectories including feedback can be solved analytically. In this
case older techniques, based on quantum Langevin equations
\cite{YamImoMac86,HauYam86,Sha87,Wis94a} can also be used. However for 
nonlinear systems, a numerical solution of the non-Markovian quantum trajectories
is the only recourse.

The simplest  system with nonlinear dynamics is the two-level atom.
Non-Markovian feedback for controlling this system was considered
earlier by two of us \cite{WanWisMil01}. We considered the simplest
form of non-Markovicity, a time delay $\tau$ in the feedback loop
\cite{GioTomVit99}
which was otherwise kept exactly as for the Markovian feedback in
Ref.~\cite{WanWis01}. We showed numerically that the time delay had an
effect  qualitatively similar to that of inefficient detection. For
the special case where the Markovian feedback would stabilize the
atom in the excited state we obtained an approximate analytical
expression for the purity (as measured by $p=2{\rm Tr}[\rho^{2}]-1$)
in the presence of a time delay. The result
for short delays, which was found numerically to be valid for quite
large delays, was
\beq \label{eq:pur}
p=1-4\gamma\tau.
\eeq
Here $\gamma$ is the decay rate for the atom. That is, the attainable
purity decreases linearly with the time delay. This appears to be
true in general for this system.

It should not be concluded from this result that non-Markovian feedback is
necessarily worse than Markovian feedback. A different paradigm for
quantum feedback has recently been developed
by Doherty {\it et al.} \cite{DohJac99,Doh00}.
It is based on an analogy with classical feedback
according to so-called ``modern control theory'' \cite{Jac93}.
Conceptually, the change is from basing the feedback directly on
 the measurement results,
to basing the feedback on an estimate of the system state. That state
estimate is of course based on the measurement results, but the extra
step usually makes the feedback non-Markovian from the point of view of the
system. That is because the best state estimate will use all previous
measurement results, not just the latest ones.

Determining the conditioned state of the quantum system from classical
measurement results is a quantum version of Bayesian
reasoning. Classical Bayesian reasoning
 updates an observer's knowledge of a system (as
described by a probability distribution over its variables) based on
new data \cite{Jac93}. For this reason, we call feedback based on a state
estimate {\em Bayesian feedback}. In classical control theory it is
common to replace Bayesian feedback with a simpler approximation to it.
For example, a linearization approximation leads to the Kalman filter,
which makes the feedback a {\em linear} functional of the observed 
current \cite{Jac93}.
The quantum version of this was explored in Refs.~\cite{DohJac99,Doh00}, and had 
previously been treated in Ref.~\cite{Bel87}.

In this paper we investigate what improvement is offered by Bayesian
feedback over Markovian feedback
for the simple problem discussed above, stabilizing an
arbitrary state of the two-level atom. We begin in Sec.~II by
discussing the
different sorts of feedback in a general context. Then we present the
specific system of interest, the two-level atom, in Sec.~III. This is
more general than that considered previously 
\cite{WanWis01,HofHesMah98,HofMahHes98,WanWisMil01} in that we include a term
in the master equation corresponding to dephasing, as caused for
instance by elastic collisions with other (background) atoms. In
Sec.~IV we present and discuss the performance of Markovian feedback
in this system. In Sec.~V we do likewise for Bayesian feedback. In
Sec.~VI we consider the prospects for approximating this Bayesian feedback
so that the feedback is a linear functional of the current.
 We conclude with a discussion in Sec.~VII.

\section{Quantum Feedback}

\subsection{Quantum Trajectories}
Quantum trajectories are the stochastic paths followed by the state
of an open
quantum system conditioned on the monitoring of its environment. In
this context, the state of the system mean the state of knowledge
of the system that an ideal observer (unlimited by computational
power) would have given the
results of the monitoring. As we
cannot assume that this monitoring will give
complete knowledge of the system, the quantum trajectory will
not be a path in Hilbert space. Rather, it will in
general be a path in the (Banach) space of state matrices $\rho$.
This path is generated by stochastic and nonlinear equation for
the conditioned state matrix, which we call a stochastic master
equation (SME).
Its classical analogue is the Kushner-Stratonovich equation for a
probability distribution \cite{Doh00}.

The system may be coupled to many independent baths, but let us assume
for simplicity that only one bath is monitored. Then we
write the (deterministic) master equation as
\beq \label{eq:SME1}
\dot{\rho} = {\cal L}{\rho} = {\cal L}_{0}\rho + {\cal D}[c]\rho,
\eeq
where the last term, described by the Lindblad \cite{Lin76}
superoperator ${\cal D}[c]\rho=c\rho c^{\dag}-\{c^{\dag}c,\rho\}/2$,
is that which is ``unraveled'' \cite{Car93} by monitoring the
relevant bath. This monitoring yields a current $I(t)$, and we denote
the state conditioned on this record ${\bf I}_{[0,t)} = \{I(s):0\leq 
s < t\}$ up to time $t$ by $\rho_{\bf I}(t)$.
The SME for this conditioned system state
$\rho_{\bf I}$ can then be written \cite{DalDziZur01}
\beq \label{eq:SMEcon1}
d\rho_{\bf I} = {\cal L}\rho_{\bf I} dt + {\cal U}\rho_{\bf I} dt,
\eeq
where
\beq \label{eq:defU1}
{\cal U}\rho \equiv (I-\bar{I})dt({\cal M}-\bar{\cal M})\rho.
\eeq
Here $I$ represents the measurement result in the infinitesimal
interval $[t,t+dt)$, which has the expected value ${\rm
E}[I]=\bar{I}$. The notation $\bar{\cal M}$, on the other hand,
represents ${\rm Tr}[{\cal M}\rho]$, where ${\cal M}$ is a
superoperator. The form of \erf{eq:defU1} guarantees two necessary
conditions: ${\rm Tr}[{\cal U}\rho]=0$, and ${\rm E}[{\cal
U}\rho]=0$. These imply that the SME preserves trace and, on average,
reproduces the master equation. In addition, ${\cal U}$ must satisfy
\beq \label{eq:proj}
\{\pi,({\cal D}[c] + {\cal U})\pi\} + dt [{\cal U}\pi ][{\cal U}\pi ] =
({\cal D}[c] + {\cal U})\pi
 \label{prop1}
\eeq
for an arbitrary rank-one projector $\pi$. This implies that, if ${\cal
D}[c]$ were the only irreversible term, the monitoring would maintain
the purity of the state.

For the case of homodyne detection we have \cite{Car93,WisMil93a}
\beq \label{eq:defM}
{\cal M}\rho = c\rho + \rho c\dg\,.
\eeq
The homodyne current $I$ is a real-valued stochastic variable satisfying
\beq \label{eq:Idtsq}
(Idt)^{2} = dt\,,
\eeq
and
\beq \label{eq:Iave}
\bar{I} = \bar{\cal M} = {\rm Tr}[\rho(c+c\dg)]\,.
\eeq
In other words,
\beq \label{eq:Idt}
I dt = {\rm Tr}[\rho_{\bf I}(c+c\dg)]dt + dW\,,
\eeq
where $dW$ is an infinitesimal Wiener increment \cite{Gar85}.

So far we have considered efficient detection. If an efficiency
$\eta<1$ is included then the conditional evolution will no longer
preserve purity. However \erf{eq:SMEcon1} still applies.
The only difference is that in the
equations for ${\cal M}$ and $I$, $c$ is replaced by $\sqrt{\eta}\,c$.
In particular, $I(t)$ becomes
\beq \label{eq:Idt2}
I dt = \sqrt{\eta}{\rm Tr}[\rho_{\bf I}(c+c\dg)]dt + dW\,.
\eeq
This is simple to understand, as the Lindbladian ${\cal D}[c]$ can be
split into $\eta{\cal D}[c] + (1-\eta){\cal D}[c]$, with only the
former being unraveled.

\subsection{Markovian Feedback}

Consider Markovian \cite{fn2} feedback of the homodyne photocurrent.
Since this current is singular and of indefinite sign, the only
possible form
of Markovian feedback is via a Hamiltonian
\beq \label{eq:Hfb1}
H_{\rm fb}(t) = {F}(t)\times I(t),
\eeq
with $F$ an Hermitian operator. 
%The factor of $\sqrt{\eta}$ is
%included so that the strength of the determinstic part of the
%Hamiltonian (proportional to $\bar{I}$) is independent of the
%efficiency.

Although $H_{\rm fb}$ at time $t$ contains the current $I$ at the
same time, it must act after the measurement. Taking this, and the
singularity of $I(t)$ into account, yields the following stochastic
equation for the conditioned system state with feedback
\cite{WisMil93b}
\bqa \label{eq:SMEcon3}
d\rho_{\bf I}&=& dt\cu{ {\cal L}_{0}\rho_{\bf I} + {\cal
D}[c]\rho_{\bf I} + {\cal
D}[F]\rho_{\bf I} -i[F,{\cal M}\rho]}
\nl{+} (I-\bar{I})dt({\cal M}' - \bar{\cal M}')\rho_{\bf I}\,.
\eqa
Here
\beq
{\cal M}'\rho \equiv {\cal M}\rho -i[F,\rho]\,.
\eeq

As noted in the introduction, the great theoretical convenience 
offered by Markovian
feedback is that it is a simple matter to remove the nonlinearity and
stochasticity in this equation by taking an ensemble average. This
replaces $I(t)$ by $\bar{I}$, yielding the Wiseman-Milburn
feedback master equation
\beq
\dot\rho = {\cal L}_{0}\rho + {\cal D}[c]\rho -i\sqrt{\eta}[F,c\rho+\rho c\dg]
+ {\cal D}[F]\rho\,.
\eeq

\subsection{Bayesian Feedback}

Following the lines sketched by Doherty et al. \cite{DohJac99}
we now consider controlling the system dynamics using a Hamiltonian
that depends not directly on the current, but rather on the observer's
state of knowledge of the system $\rho_{\bf I}$. By definition there
is nothing better with which to control the system. We thus
have in general
\beq \label{eq:Hfb2}
H_{\rm fb} = {F}(t,\rho_{\bf I})\,.
\eeq
It is an odd fact about Bayesian feedback that,
although strictly it is non-Markovian, if the experimenter
controlling the system has perfect knowledge of the system dynamics,
then the system state actually does obey a Markovian equation, namely
\beq \label{eq:SMEcon4}
d\rho_{\bf I} = dt\left[{\cal L}+ {\cal U}\right]\rho_{\bf I}
-i[{F}(t,\rho_{\bf
I}),\rho_{\bf I}]\,.
\eeq
However it is not possible to average over the stochasticity to obtain
a master equation. This reveals the underlying non-Markovicity.

The presence of a nonlinear stochastic Markovian equation for the
conditioned system state is an artifact of the assumption of perfect
knowledge of the system dynamics. In reality the system dynamics
would not be known perfectly, and the experimenter's estimate 
$\check\rho_{\bf I}$ of the
system state $\rho_{\bf I}$ would be governed by an equation different from
\erf{eq:SMEcon1}, namely
\beq \label{eq:SMEcon5}
d\check\rho_{\bf I} = \tilde{\cal L}_{\check\rho}\check\rho_{\bf I}
dt +
\tilde{\cal U}_{\check\rho}\check\rho_{\bf I} dt,
\eeq
where
\beq \label{eq:defU2}
\tilde{\cal U}_{\check\rho} \check\rho \equiv
(I-\bar{I}_{\check\rho})dt
({\cal M}-\bar{\cal M}_{\check\rho})\check\rho.
\eeq
Here $\tilde{\cal L}_{\check\rho}$ is an approximation to ${\cal L}$.
The approximation may be necessary due to lack of information, or it
may be convenient to allow a simpler treatment of the system.  This
approximation may depend on the estimated system state
$\check{\rho}_{\bf I}$. The stochastic unraveling
superoperator $\tilde{\cal U}$ may also be approximated for reasons
such as these, with ${\cal M}$ replaced by $\tilde{\cal M}$.
However in \erf{eq:defU2} we have shown it as approximate
for a necessary reason, namely that in general it depends upon an
estimate of
$\rho$, $\check\rho$, in order to evaluate $\bar{I}$ and $\bar{\cal
M}$.

Linearization of dynamics is a good example
of a convenient approximation. It is typically applied to systems
with infinite dimensional Hilbert spaces, corresponding to a classical
phase space. Under linear dynamics of such a system, the conditioned
state of the system will tend towards a Gaussian state. For a system
with $N$ co-ordinates ($2N$ phase-space variable), the
state $\check\rho_{\bf I}$ is describable by $2 N^{2} + 3
N$ variables, recording the covariance matrix and the means. This
compares with of order $D^{2N}$ real numbers required to record $\rho_{\bf I}$,
where $D$ is an approximation to infinity. Moreover, the equation for
the co-variance matrix is deterministic, and that for the means is
linear. This is what leads to the Kalman filter, where the
feedback is a {\em linear} functional of the observed current 
\cite{Jac93}.

If the experimenter's best estimate of the system is $\check\rho_{\bf
I}$ then with the feedback included this estimate would still obey a
Markovian equation, namely
\beq \label{eq:SMEcon6}
d\check\rho_{\bf I} = dt\left[\tilde{\cal L}_{\check\rho}+ \tilde{\cal
U}_{\check\rho}\right]\check\rho_{\bf I}
 -i[ {F}(t,\check\rho_{\bf
I}),\check\rho_{\bf I}].
\eeq
However a second, more diligent, observer would
use the full knowledge of the system dynamics to obtain the system
state $\rho_{\bf I}$. This would obey the
stochastic master equation
\beq \label{eq:SMEcon7}
d\rho_{\bf I} = dt\left[{\cal L}+ {\cal U}\right]\rho_{\bf I}
-i[ {F}(t,\check\rho_{\bf
I}),\rho_{\bf I}]\,.
\eeq
Note that this is not a Markovian equation for $\rho_{\bf I}$, because
the feedback depends on the estimate $\check\rho_{\bf I}$. The two
equations together are Markovian, and in control theory language this
would be considered an example of Markovian control. However from the
usual perspective of quantum mechanics, where the ``system'' is the
quantum system, not the quantum system plus control loop, this is an
example of non-Markovian feedback control.

\section{The System}

The simplest nonlinear system to consider is an atom, with two relevant levels
$\{\ket{g},\ket{e}\}$ and lowering operator
$\sigma=\ket{g}\bra{e}$. Let the decay rate be unity, and let
it be driven by a resonant classical driving field with Rabi
frequency $2\alpha$. %This is as shown in Fig.~\ref{fig:diag},
%where for the moment
%we are omitting feedback by setting $\lambda=0$.
Furthermore, let us add dephasing of the atomic dipole at rate 
$\Gamma$.

\subsection{The Master Equation}

The evolution of this  system is described by the master
equation
\beq \label{eq:SME3}
\dot{\rho} =  {\cal D}[\sigma]\rho - i\alpha [\sigma_{y},\rho]
+\Gamma{\cal D}[\sigma_{z}] \equiv {\cal L}\rho\,.
\eeq
In this master equation
 we have chosen to define the $\sigma_{x}=\sigma+\sigma\dg$ and
$\sigma_{y}=i\sigma-i\sigma\dg$ quadratures of the atomic dipole
relative to the driving field. The effect of driving is to rotate the
atom in Bloch space around the $y$-axis. The state of the atom in
Bloch space is described by the three-vector $(x,y,z)$. It is related
to the state matrix $\rho$ by
\beq \label{eq:rhoBlo}
\rho = \frac{1}{2}\ro{I + x\sigma_{x}+y \sigma_{y}+z\sigma_{z}}.
\eeq

\begin{figure}[tbp]
\includegraphics[width=0.45\textwidth]{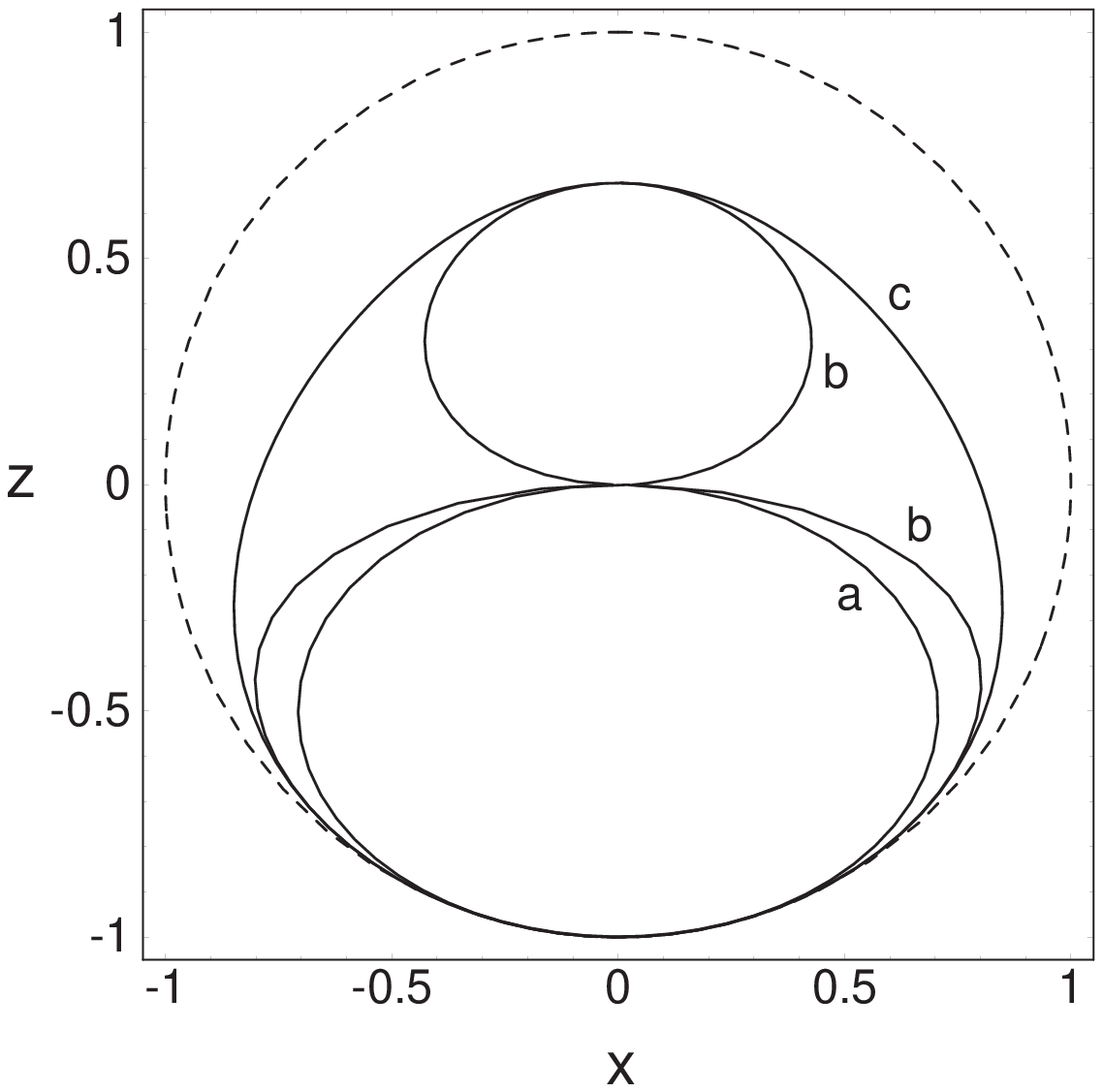}
\caption{\narrowtext Locus of the ensemble average
solutions to the Bloch equations with detector efficiency
$\eta = 0.8$ and dephasing rate $\Gamma=0$ under various conditions:
(a) no feedback (driving only); (b) Markovian feedback;
(c) Bayesian feedback. The dashed line is the surface of the Bloch 
sphere which is stabilizable for $\eta=1$ by Bayesian feedback and 
(except for the equatorial points) by Markovian feedback.}
	\protect\label{fig:eta}
\end{figure}

\begin{figure}[tbp]
\includegraphics[width=0.45\textwidth]{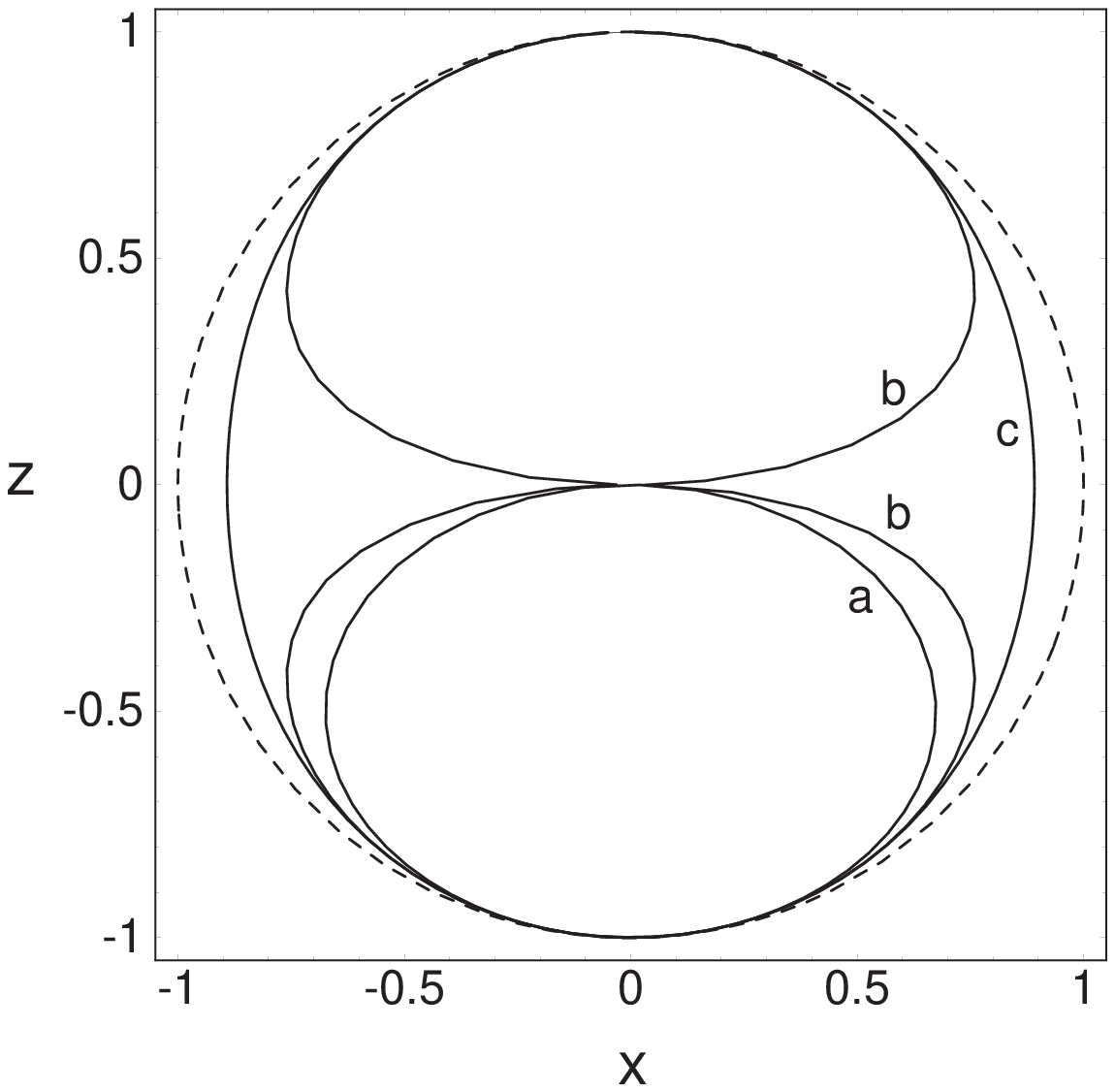}
\caption{\narrowtext Locus of the ensemble average
solutions to the Bloch equations with detector efficiency
$\eta = 1$ and dephasing rate $\Gamma=1/20$ under various conditions:
(a) no feedback (driving only); (b) Markovian feedback;
(c) Bayesian feedback. The dashed line is the surface of the Bloch 
sphere which is stabilizable for $\Gamma=0$ by Bayesian feedback and 
(except for the equatorial points) by Markovian feedback.}
	\protect\label{fig:Gamma}
\end{figure}

It is easy to show 
that the stationary solution of the master
equation (\ref{eq:SME3}) is
\bqa
{x}_{\rm ss}&=&\frac{-4\alpha }{ (1+2\Gamma)+8\alpha^2},
\label{eq:xss1}\\
{y}_{\rm ss}&=&0,
\label{eq:yss1}\\
{z}_{\rm ss}&=&\frac{- (1+2\Gamma)}{ (1+2\Gamma)
+8\alpha^2}.
\label{eq:zss1}
\eqa
For $\Gamma$ fixed,
this is a family of solutions parameterized by the
driving strength $\alpha \in (-\infty,\infty)$. All members of the
family are in the $x$--$z$ plane on the Bloch sphere. Thus for this
purpose we can reparametrize the relevant states using $r$ and
$\theta$ by
\bqa
x &=& r\sin \theta, \label{eq:rsin}\\
z &=& r\cos \theta, \label{eq:rcos}
\eqa
where $\theta \in [-\pi,\pi]$. Since
\beq \label{eq:defpur}
p = 2{\rm Tr}[\rho^{2}] - 1 =  x^{2}+y^{2}+z^{2},
\eeq
is a measure of the purity of the Bloch sphere,
$r=\sqrt{x^{2}+z^{2}}$,
the distance from
the centre of the sphere, is also a measure
of purity. Pure states correspond to $r=1$ and maximally mixed
states to $r=0$.
The stationary states we can reach by driving
the atom are limited, and generally far from pure \cite{WanWis01}.
In particular, they are confined to the lower half of the Bloch
sphere, as shown in Figs.~\ref{fig:eta} and \ref{fig:Gamma}.

\subsection{Homodyne Measurement}

Now consider subjecting the atom to homodyne detection. 
%As shown in
%Fig.~\ref{fig:diag}, 
We assume that all of the fluorescence of the
atom is collected and turned into a beam. This could be achieved in 
principle by placing the atom at the focus of a parabolic mirror, but 
in practice it is more likely to be 
achievable in a cavity QED setting \cite{Tur98}, with the atom strongly coupled 
($g$) to 
to a single cavity mode, which is
strongly damped ($\kappa$). Then the 
combined system acts like an effective two-level atom, and the output beam
of the cavity is effectively the spontaneous emission of the atom, 
with the rate (which we have defined as unity) being $O(\kappa/g^{2})$.
%(represented in
%Fig.~\ref{fig:diag} by
%placing the atom at the focus of a mirror). 
Under homodyne measurement of the  $x$
quadrature of the output field, the conditioned state
will continue to be confined to the $x$--$z$ plane. 
In this case the homodyne
photocurrent
is given by
\beq \label{eq:Iatom}
I(t)dt = \sqrt{\eta}{\rm Tr}[\rho_{\bf I}\sigma_{x}]dt + dW(t)\,,
\eeq
and the measurement superoperator by
\beq
{\cal M}\rho = \sqrt{\eta}(\sigma \rho + \rho\sigma\dg)\,.
\eeq
The conditioning SME is thus
\beq
d\rho_{\bf I} = {\cal L}\rho_{\bf I} dt + (I-\bar{I})dt({\cal
M}-\bar{\cal
M})\rho_{\bf I}.
\eeq

\section{Markovian Feedback} \label{sec:mf}

Markovian feedback in this system has been considered before 
\cite{WanWis01}, except
for the effect of dephasing $\Gamma$. This can be treated by the same
techniques, so our presentation here will be brief. The aim of this
feedback scheme, and indeed all feedback schemes considered in this
paper, is to make the stationary state of the atom as close as
possible to a pure state $\ket{\theta_{0}}$, defined by
\beq
\ket{\theta_{0}} = \cos\frac{\theta_{0}}{2} \ket{e} + \sin
\frac{\theta_{0}}{2} \ket{g} .
\eeq
Here $\theta_{0}$ is a given parameter in $[-\pi,\pi)$. The state
$\ket{\theta_{0}}$ is a state with $r$ and $\theta$, as
defined above, given by $r=1$ and $\theta=\theta_{0}$.

Since the desired state is in the $y=0$ plane, control of the atomic
state can be effected by a feedback Hamiltonian proportional
to $\sigma_{y}$. For Markovian feedback we have
\beq
H_{\rm fb} = I(t) \lambda \sigma_{y}/\sqrt{\eta},
\eeq
where $\lambda$ is the feedback parameter. Since the driving
Hamiltonian is $\alpha \sigma_{y}$, this feedback  is
physically
realized simply by  modulation of the driving.

The deterministic master equation including feedback is, in the
Lindblad form,
\beq \label{eq:SMEfb1}
\dot{\rho} = -i[\alpha\sigma_{y},\rho]+{\cal
D}[\sigma-i\lambda\sigma_{y}]\rho +
\frac{\lambda^{2}}{\eta}{\cal D}[\sigma_{y}]\rho
+\Gamma{\cal D}[\sigma_{z}].
\eeq
We do not know {\em a priori} what values of $\lambda$ and $\alpha$ to
choose to give the best results. Hence we
simply solve for the stationary matrix in terms of $\alpha$ and
$\lambda$. Using the Bloch representation we find
\bqa
{x}_{\rm ss}&=&{-4\alpha(1+2\lambda)}/D,
\label{eq:xss2}\\
{y}_{\rm ss}&=&0 ,
\label{eq:yss2}\\
{z}_{\rm ss}&=&-(1+2\lambda)
\left(1+4\lambda+2\Gamma+4{\lambda^2}/{\eta}
\right)\Big/D,
\label{eq:zss2}
\eqa
where
\bqa \label{eq:D}
D =
8\alpha^{2}&+&\left(1+4\lambda+2\Gamma
+4 {\lambda^{2}}/{\eta}\right)
\nn\\
&&\times
\left(1+2\lambda+2{\lambda^{2}}/{\eta}\right)\,.
\eqa

The ``best results'' for the
feedback system is achieved by maximizing the radius $r$ in
\erf{eq:rsin} and
\erf{eq:rcos} for each $\theta_{0}$. From these two equations we have
\beq
\tan\theta = x_{\rm ss}/z_{\rm ss}.
\eeq
From \erfs{eq:xss2}{eq:zss2} we can immediately find the
desired driving in terms of $\lambda$ and $\theta_{0}$ as
\beq\label{alinlam}
\alpha =
\left(1/4+\lambda+\Gamma/2+{\lambda^2}/{\eta}\right) \tan\theta_{0}. 
\label{alpha}
\eeq
The aim is then, for each $\theta_{0}$, to find the feedback
$\lambda$ which
maximizes
\bqa
r_{\rm ss} &=& \sqrt{x_{\rm ss}^{2}+z_{\rm ss}^{2}} \\
&=& \frac{(1+2\lambda)\cos\theta_{0}}
{1+2\lambda+2\lambda^{2}/\eta + (\Gamma-1/2)\sin^{2}\theta_{0}}
\eqa

We find that
\beq
{\rm max} \, r_{\rm ss} = r_{0}\,,
\eeq
where $r_{0}$ is the
solution of
\bqa \label{defr0}
0 &=& r_{0}^{2}\left[ \left(1 - \eta +\cos^{2}\theta_{0}\right)/2
+ \Gamma \sin^{2}\theta_{0} \right]
\nl{+} r_{0}(1-\eta)\cos\theta_{0}  -  \eta (\cos^{2}\theta_{0} )/2\,.
\eqa
This maximum is achieved for
\beq
\lambda = -\frac{\eta}{2}\left(1+r_{0}^{-1}\cos\theta_{0}\right).
\eeq
Note that for $\eta \neq 1$,
this optimal $\lambda$, and the resultant $r_{0}$, were only
found numerically in previous work \cite{WanWis01}.  The analytical
results here, which also include $\Gamma \neq 0$, are new.

The curve resulting from \erf{defr0} is shown in Fig.~\ref{fig:eta}
and Fig.~\ref{fig:Gamma} for different parameters. For prefect
conditions ($\eta=1$ and
$\Gamma=0$) it is possible to stabilize any state $\ket{\theta_{0}}$
except those on the equator (see Sec.~\ref{perfcon} below). Under imperfect
conditions,
the maximum purity $r_{\rm ss}$ decreases, with a gap opening up at
the equator. For inefficient detection, the purity of the optimal
states in the upper	half of	the	Bloch sphere is	affected much more
than those
in the lower half, whereas the two halves remain symmetrical for
non-zero phase diffusion. This is explicable as follows. In the limit $\eta
\to 0$ (no detection) the feedback cannot be effective, so the locus
of states must reduce smoothly as $\eta \to 0$ to the no-feedback
result also shown in Fig.~\ref{fig:eta}. By contrast, as $\Gamma$
increases there is no necessity that the no-feedback result should be
recovered, and moreover the phase diffusion term $\Gamma {\cal
D}[\sigma_{z}]$ is unchanged by reflection about the equator
($\sigma_{z} \to -\sigma_{z}$). A number of cases of interest now need
consideration.

\subsection{Perfect Conditions} \label{perfcon}

In the case $\eta=1$, $\Gamma=0$ the above parameters simplify greatly.
We find, in agreement with Ref.~\cite{WanWis01},
\bqa
\alpha &=& (\cos\theta_{0}\sin\theta_{0})/4\,, \\
\lambda &=& -(1+\cos\theta_{0})/2\,.
\eqa
With these parameters any $\ket{\theta_{0}}$ can be stabilized,
except $|\theta_{0}|=\pi/2$ (that is, on the equator of the Bloch
sphere). This is clear from the parameters, since
the values of $\alpha$ and $\lambda$ are the same for
$\theta_{0}=\pi/2$ and $\theta_{0}=-\pi/2$. For a two-level system the
same master equation cannot have two different stable stationary
states.
Thus the equatorial states cannot be stabilized by Markovian feedback
even under perfect conditions of  efficient detection and no
dephasing.

\subsection{Stabilizing the Excited State}

Another case where the parameters (and the purity)
have simple expressions is for
$\theta_{0}=0$; that is, trying to stabilize near the excited state.
For this we desire $x_{\rm ss}=0$ so $\alpha=0$. We find from
\erf{defr0} that
\beq \label{esp}
z_{\rm ss} = {\rm E}[r] = \frac{\eta}{2-\eta},
\eeq
for $\lambda = -1$. Another simple case is  stabilizing the ground state.
This is of course always possible to do perfectly, simple by
turning the feedback and driving off.

\subsection{Stabilizing an Equatorial State}

A final case where the purity can be found analytically is for
$\theta_{0}=\pm \pi/2$. That is, trying to stabilize an equatorial
state. Markovian feedback cannot achieve this at all. From
\erf{eq:xss2} and \erf{eq:zss2}, if $z_{\rm ss}=0$ then necessarily
$x_{\rm ss}=0$  also. The stochastic conditioned dynamics that underly 
this were explored in Ref.~\cite{WanWis01}.

\section{Bayesian Feedback}

Because Bayesian feedback is based on knowledge of
the conditioned state
$\rho_{\bf I}$, we need to examine its evolution in \erf{eq:SMEcon1}
in more detail. As noted above, the state is confined to the $y=0$
plane, so it is very convenient to  write the evolution
 in terms of $r$ and $\theta$ as defined in \erfs{eq:rsin}{eq:rcos}. 
 Using the \ito stochastic calculus \cite{Gar85}, the result is
\bqa
dr_{\bf I} &=&
\left\{-r_{\bf I}\left(1+\cos^{2}\theta_{\bf I}\right)/2
-\Gamma r_{\bf I}\sin^{2}\theta_{\bf I}
-\cos\theta_{\bf I}\right.
\nonumber\\
&&+\left.
\frac{\eta}{2} \left[
\frac{\cos^{2}\theta_{\bf I}}{r_{\bf I}}
+2\cos\theta_{\bf I}+r_{\bf I}
\right]\right.
\nonumber\\
&&+\left.\sqrt{\eta}\left[\sin\theta_{\bf I}\left(1-r^{2}_{\bf
I}\right)
\right][I(t)-\sqrt{\eta}\,r_{\bf I}\sin\theta_{\bf I}]\right\}dt
\,,\nonumber\\
\label{eq:dr1} \\
d\theta_{\bf I} &=&  \left\{
\left(\frac{1}{2}-\Gamma\right)\sin\theta_{\bf I}\cos\theta_{\bf I}
+2\alpha+\frac{\sin\theta_{\bf I}}{r_{\bf I}}\right.
\nonumber\\
&&+\left.
\eta \left[
\sin\theta_{\bf I}\left(r_{\bf I}+\cos\theta_{\bf I}\right)
\left(1-\frac{1}{r_{\bf I}^{2}}\right)\right]\right.
\nonumber\\
&&+\left.\sqrt{\eta}\left[1+\frac{\cos\theta_{\bf I}}{r_{\bf
I}}\right]
[I(t)-\sqrt{\eta}\,r_{\bf I}\sin\theta_{\bf I}]\right\}dt
\,.\nonumber\\
\label{eq:dtheta1}
\eqa

For perfect state-estimation, the experimenter knows these values of
$r_{\bf I}$ and $\theta_{\bf I}$ from the measurement record $\bfi$.
Now we
wish to
add feedback, the aim of which is to stabilize the state of the system
to be as close as possible to a given pure state
$\ket{\theta_{0}}$. We again consider  feedback by modulation of
the driving Hamiltonian, where the modulation can depend in an
arbitrary way upon $\rho_{\bf I}$ (that is, $r_{\bf
I}$ and $\theta_{\bf I}$). This can change $\theta_{\bf I}$ but not
$r_{\bf I}$.  To
maximize the closeness to the state $\ket{\theta_{0}}$ we wish to
force $\theta_{\bf I}$ to equal $\theta_{0}$. This is achieved by the
feedback Hamiltonian
\beq \label{eq:Hfb3}
  H_{\rm fb} =  F(\rho_{\bf I}) = \lim_{\beta\to\infty}
-\beta \sigma_{y} (\theta_{\bf I}-\theta_{0})\,. 
\label{Hfb2}
\eeq
This adds the term
\beq \label{eq:fb}
\lim_{\beta\to\infty} -2\beta(\theta_{\bf I}-\theta_{0})dt\,,
\eeq
to $d\theta_{\bf I}$ in \erf{eq:dtheta1}.

Clearly with the limit
$\beta\to\infty$  this term will suppress all fluctuations in
$\theta_{\bf
I}$ and force it to take the value $\theta_{0}$. The SME for the
system then reduces to a single equation for $r_{\bf I}$, found by
substituting $\theta_{\bf I}=\theta_{0}$ in \erf{eq:dr1}:
\beq \label{eq:dr2}
dr_{\bf I}=  A(r_{\bf I})dt + \sqrt{B(r_{\bf I})} dW(t),
\eeq
where
\bqa
A(r) &=& -r\left(1+\cos^{2}\theta_{0}\right)/2
-\Gamma r\sin^{2}\theta_{0}
-\cos\theta_{0},
\nonumber\\
&&+ \frac{\eta}{2}\left[
\frac{\cos^{2}\theta_{0}}{r}
+2\cos\theta_{0}+r
\right] \\
B(r) &=& {\eta} \sin^{2}\theta_{0}\left(1-r^{2}\right)^{2}.
\eqa
Here we are using $dW$ for 
$I-\sqrt{\eta}\,r_{\bf I}\sin\theta_{\bf I}$.

This stochastic differential equation is equivalent to the
following Fokker-Planck equation for the probability 
$P(r) = {\rm Prob}[r_{\bf I}=r]$
\beq
\dot{P}(r) = \sq{-\dbd{r} A(r) + \frac12 \frac{\partial^{2}}{(\partial r)^{2}}
B(r)}P(r)\,.
\eeq
It is then easy to show \cite{Gar85} 
that the stationary mean of $r_{\bf I}$ is
\beq \label{rssGar}
r_{\rm ss} = \frac{\int_{0}^{1} rdr
C(r)\exp\sq{2\int_{0}^{r}dr' {A(r')}{C(r')}}   }
{\int_{0}^{1} dr
C(r)\exp\sq{2\int_{0}^{r}dr' {A(r')}{C(r')}}   },
\eeq
where $C(r)=1/B(r)$. This will clearly depend upon $\theta_{0}$.

These integrals can be easily solved numerically, and the results
 are shown in Fig.~\ref{fig:eta}
and Fig.~\ref{fig:Gamma}. Under perfect conditions, any state
$\ket{\theta_{0}}$ can be stabilized perfectly, as discussed below.
 For $\eta < 1$ or
$\Gamma > 0$ the purity decreases, in a qualitatively similar way
to Markovian feedback. However, the purity for Bayesian feedback is
better than for Markovian feedback for almost all $\theta_{0}$, and
is never worse.

\subsection{Near-Perfect Conditions} \label{notelin}

It is interesting to consider the case of near-perfect
conditions, where $r_{\rm ss} \simeq 1$. This requires $\eta \simeq
1$ and $\Gamma \ll 1$. In this case, $r_{\bf I}$
cannot typically wander far from $r_{\rm ss}$, since it is bounded
above by unity. This suggest that it may be possible to linearize
\erf{eq:dr2}, because the fluctuations are small. Assuming $\Gamma \ll
1$  and setting $\eta = 1-\epsilon$ with $\epsilon \ll 1$, we get
\beq
A(r) \simeq - \cos^{2}\theta_{0} \times (r - r_{0})\,.
\eeq
Here $r_{0}$ is the solution of \erf{defr0} for $\Gamma, \epsilon \ll 1$,
namely
\beq
r_{0} \simeq 1- \epsilon(1+1/\cos\theta_{0}) 
- \Gamma\tan^{2}\theta_{0}\,.
\eeq
Clearly this argument only works for $\theta_{0} \neq \pm \pi/2$.

\begin{figure}[tbp]
\includegraphics[width=0.45\textwidth]{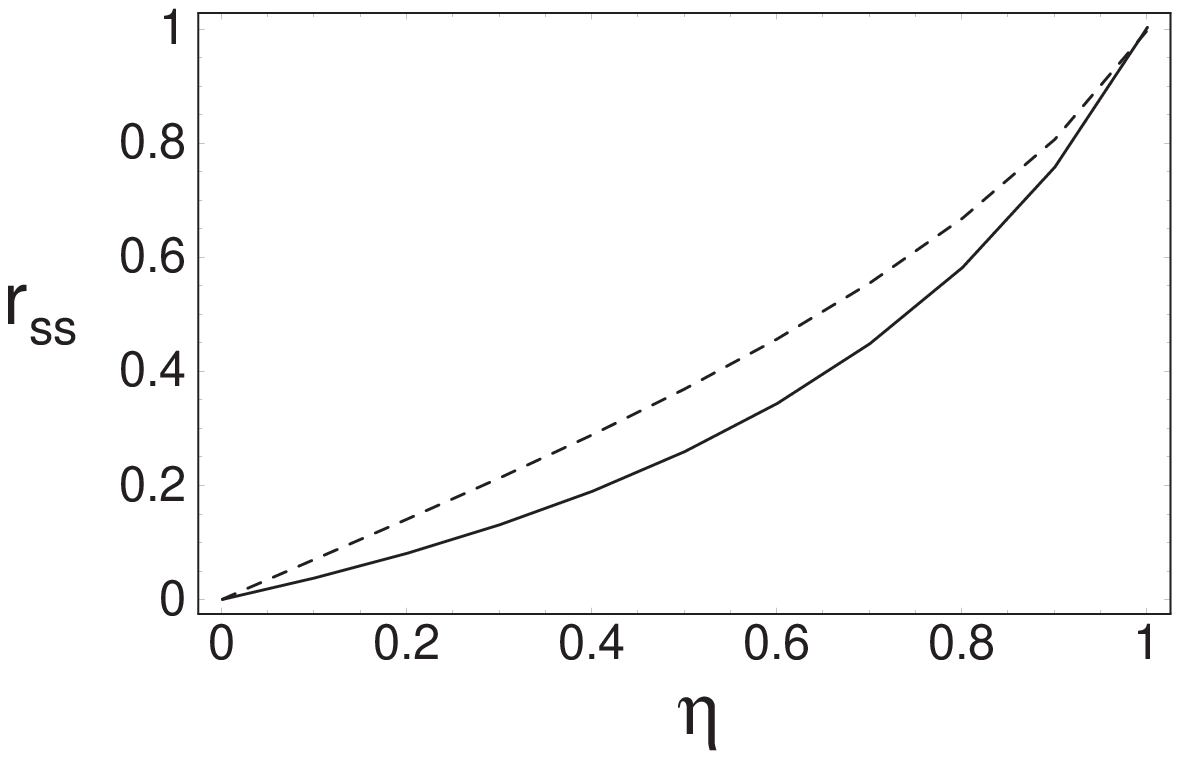}
\caption{\narrowtext Purity $r_{\rm ss}$ achievable for Markovian
(solid)
and Bayesian (dashed)  feedback as a function of $\eta$ for
$\theta_{0}=\pi/4$ and $\Gamma=0$. Note that they perform identically 
for $\eta \simeq 1$.}
	\protect\label{fig:etadec}
\end{figure}

Now it turns out that it is not valid to approximate $B(r)$ by a
constant $B(r_{0})$ because it varies rapidly when $r$ is close to one.
However, this is actually irrelevant, because as long as $A(r)$ can be
approximated by a linear function of $r$ plus a constant, the equation
for the expectation value of $r_{\bf I}$ is
\beq
\frac{d\;}{dt}{\rm E}[r_{\bf I}] = A({\rm E}[r_{\bf I}]).
\eeq
In this case, it is clear that $r_{\rm ss} = r_{0}$. That is, Bayesian
feedback can offer no improvement over Markovian feedback for the case
of near-perfect conditions. This is evident in Fig.~\ref{fig:etadec}.

\subsection{Stabilizing the Excited State}

It will have been noted from Figs.~\ref{fig:eta} and \ref{fig:Gamma} 
that Bayesian feedback is also no more effective at
stabilizing a state near the excited state $\ket{0}$ than Markovian
feedback. This can be proven analytically. With $\theta_{0}=0$ we have
from \erf{eq:dr2}
\beq
\frac{dr}{dt} = -r - 1 + \frac{\eta}{2}\ro{\frac{1}{r}+2+r},
\eeq
which is independent of $I(t)$ (and of $\Gamma$) and has the stationary
solution
(\ref{esp}). That is, here is another case where Bayesian feedback
offers no improvements over Markovian feedback.

\subsection{Stabilizing an Equatorial State}

One case where Bayesian feedback clearly has an advantage over
Markovian feedback is for stabilizing an equatorial state.
At first sight this seems in contradiction with \erf{eq:dr2},
which for $\theta_{0}=\pi/2$ becomes
\beq \label{eq:dr4}
dr_{\bf I} = -[\Gamma +  (1-\eta)/2]r_{\bf I}dt +
\sqrt{\eta}(1-r_{\bf
I}^{2})dW.
\eeq
This has an expectation value that decays to zero, the same as in the
Markovian case. However, this
equation also allows $r_{\bf I}$ to become negative, which
invalidates
the basis for the equation, namely that $\theta_{\bf I}$ is fixed at
$\theta_{0}$ so that $r_{\bf I}$ is positive and represents the
purity. If $r_{\bf I}$ becomes negative then this indicates that
$\theta_{\bf I}$ has switched from $\pi/2$ to $-\pi/2$ say. Bayesian
feedback could then correct this. This can be treated with the above
equation (\ref{eq:dr4}) if we assume that whenever $r_{\bf I}$
becomes
negative it is made positive (with its magnitude unaltered). This
sort of assumption has already been used in \erf{rssGar} in setting
the lower limits of the integrals to zero. In the limit of large
$\Gamma$ or small $\eta$, where $r_{\bf I}$ will tend to be small,
we can approximate the coefficient of the noise term in
\erf{eq:dr4} by  $\sqrt{\eta}$. Then \erf{rssGar} can be solved
analytically to yield
\beq
r_{\rm ss} \simeq \sqrt\frac{\eta}
{[\Gamma +  (1-\eta)/2]\pi}.
\eeq

\section{Linearized Bayesian Feedback?}

The Bayesian feedback
 of the previous section was unrealistically perfect in two
aspects. First, we allowed for infinitely strong feedback.
However, this is
only fair for comparison with Markovian feedback since it also allows
for infinitely strong feedback, since the current $I(t)$ in the
Markovian feedback Hamiltonian is a singular function of time.
Second, we assumed that the state estimation was perfect. That is, we
assumed that the experimenter could solve the nonlinear stochastic
Bloch
equations in real time to obtain $\rho_{\bf I}$ and hence $\theta_{\bf
I}$. This is a much more demanding task than for Markovian feedback,
where the current is fed-back without any processing. It would thus
be of interest to see how well Bayesian feedback performs if the
processing is reduced to a level more comparable with that required
in Markovian feedback. Specifically, feeding back a linear functional
of the current would be well comparable.

What we desire is a systematic way of deriving an appropriate
linearized Bayesian
feedback of this sort. An obvious approach is to linearize the stochastic
equations of motion for the state vector parameters 
$r_{\bf I}$ and $\theta_{\bf I}$. Assuming
that this feedback does approximate the full Bayesian feedback, the
linearization for $\theta_{\bf I}$ should be done about $\theta_{0}$.
It then follows that the linearization for $r_{\bf I}$ should be done
about the deterministic fixed point of \erf{eq:dr2}. That is, the
point $r_{0}$ satisfying $A(r_{0})=0$. But this $r_{0}$ turns out to
be exactly the same as the $r_{0}$ defined by \erf{defr0}. That is, it
seems that we should linearize about the point that is the stationary
solution of the Markovian feedback master equation.

If the
linearization of the dynamics were valid, then the
variables $\check{r}_{\bf I}, \check{\theta}_{\bf I}$ with linear dynamics
would be
good approximations to the exact variables $r_{\bf I}$. Let us 
consider the best case scenario where the feedback would be
a good enough approximation to the full Bayesian
feedback for  fluctuations in
$\theta_{\bf I}$ to be completely suppressed. Then
$\check{r}_{\bf I}$ would obey a linearized version of \erf{eq:dr2},
and the stationary mean solution would give $r_{\rm ss}$. But, as
already noted in Sec.~\ref{notelin}, if the drift term
$A(\check{r}_{\bf I})$ consists of a constant and a linear term, then
the stationary mean solution is the fixed point $r_{0}$. In other
words, ${\rm E}[\check{r}]_{\rm ss} = r_{0}$.

This fact bodes ill for linearized Bayesian feedback. If the
linearization were valid then $r_{\bf I} \simeq \check r_{\bf
I}$ and so $r_{\rm ss} \simeq r_{0}$. That is, the linearized
Bayesian feedback would do no better than Markovian feedback. If the
linearization were not valid, then there would be no reason to expect the
linearized algorithm to work at all. It is quite possible that by
fluke there is some linear functional of the current $I(t)$ that
would give a better result than Markovian feedback. However that is of
no great conceptual significance, if the linear functional is not
derived from an approximation to the Bayesian theory.

The linearized Bayesian feedback described above is not based on a
Kalman filter, because the variables $r$ and $\theta$ have no
correspondence with classical phase space variables. In particular,
$r$ is itself a measure of purity, and here obeys a (linearized)
stochastic equation. In the Kalman filter, the purity is
determined by the covariance matrix, which obeys a deterministic
equation. It might therefore be thought that a better approach to
linearizing the Bayesian feedback would be to approximate the surface
of the Bloch sphere by a plane, thereby creating an analogue to the
classical phase plane. Since the Bloch sphere dynamics are confined to
the transverse plane $y=0$, the tangential plane reduces to a line,
parametrized by $\theta$ in the neighbourhood of $\theta_{0}$.

In this alternative approach, the
linearization would then be based upon describing the state of the
atom by a Gaussian distribution $P(\theta)$ of states $\ket{\theta}$, 
localized about
$\theta_{0}$. Averaging over this distribution would give a purity
\beq
r = {\rm E}[\exp\ro{i(\theta-\theta_{0})}] \simeq 1 - {\rm
E}[(\theta-\theta_{0})^{2}]/2.
\eeq
It is possible to obtain an equation for $P(\theta)$ by considering 
fictitious noises (corresponding to hypothetical measurements of the 
undetected fluorescence and of the bath causing the dephasing), 
and then averaging over them. 
However to obtain a linear feedback algorithm, 
the resulting equation must be linearized, thus yielding a 
Gaussian solution for $\tilde{P}(\theta)$ with constant variance.
This would be possible only if the variance is much less than unity. 
That is, this approach could only work if $r$ were close to one.
However, we have already seen in Sec.\ref{notelin} that in this
regime, the full Bayesian feedback is no better than Markovian
feedback, so the linearized Bayesian feedback cannot do better either.
In fact, in this limit, it can be shown that direct 
linearization of the Bayesian feedback 
 reproduces the Markovian feedback.

\section{Conclusions}

In this paper we have contrasted two different approaches to quantum 
control, Markovian feedback (where the current is fed-back with no 
filtering) and Bayesian feedback (where the feedback is based upon an 
estimate of the  state). We have applied 
them to the problem of stabilizing the quantum state 
of the simplest nonlinear system, a two-level atom, to be near an 
arbitrarily chosen pure 
state $\ket{\theta_{0}}$. Due to the simplicity of our system, we are 
able to obtain all of our results without 
 numerical stochastic simulations, as required in previous work on Bayesian feedback in 
nonlinear systems \cite{Doh00}.

Unsurprisingly, Bayesian feedback never performs worse than 
Markovian feedback.
For close to ideal conditions (small atomic dephasing, and detection 
efficiency close to one),  Bayesian feedback performs 
identically to Markovian feedback, except for $|\theta_{0}| \simeq \pi/2$. In 
less ideal situations, it performs better for almost all values of 
$\theta_{0}$. However, Bayesian feedback 
is far more demanding experimentally than Markovian feedback. That 
is because it relies upon the real-time solution of nonlinear 
stochastic differential equations, namely those that determine the 
state-estimate. 

For Markovian feedback, the effect of  imperfections (such as a time 
delay in the feedback loop) have been studies \cite{WanWisMil01} and 
they are not disastrous if they are small. In the present study we 
have not considered the effect of imperfections in Bayesian 
feedback, and it is not clear how disastrous such inevitable 
imperfections would be. As a partial attempt  to this question, we have 
considered replacing the full Bayesian feedback with a linearized 
version. This would yield a feedback signal which is a linear 
functional of the feedback current, and so would also be 
experimentally more reasonable and comparable to Markovian feedback. 
Unfortunately we find that any systematic approach to such a 
linearization results in a feedback algorithm that would be expected 
to perform {\em worse} than Markovian feedback in general.

Two approaches to linearization for the two-level atom were considered. The 
first is based on treating 
the parameters $(r,\theta)$ of the state matrix $\rho$ as the objects 
to be controlled. The second describes $\rho$ as a 
narrow Gaussian mixture of state vectors $\{\ket{\theta}\}_{\theta}$ 
about $\theta_{0}$. In effect, it  seeks to control the 
hypothetical ``true state'' $\ket{\theta}$. Both of these approaches 
to quantum control were considered by 
Doherty {\em et 
al.} \cite{Doh00}, but not as paths to a linear feedback 
algorithm. Indeed, the first approach (which they actually discuss 
last) is described by them as ``necessarily nonlinear'', 
although it can of course be linear in some cases. 
The second approach they note is potentially ill-defined and in any case 
``sub-optimal''.

For the purposes of developing a linearized quantum feedback 
algorithm, Doherty {\em et al.} consider only one 
approach to quantum control. This is the one (the first they discuss) 
based on describing the quantum system by 
a quasi-probability distribution on classical phase space. 
This description for quantum systems is naturally linearized to yield 
 the Kalman filter, as discussed in Sec.~II~C. 
 The negative results obtained here for non-Kalman linearized feedback 
suggests that perhaps good linearized quantum feedback control 
algorithms exist only for quantum systems whose state can be well 
described by a classical phase space distribution. This of course 
rules out the two-level atom and other ``deep quantum'' systems.

To conclude, if there were no restrictions placed upon an 
experimenter's processing 
ability or knowledge of relevant parameters 
then Bayesian feedback would be optimal by definition. 
Moreover, we have shown that in 
most parameter regimes, it does strictly better than Markovian feedback in 
stabilizing the state of a two-level atom. However, for nonlinear 
systems (such as the atom), Bayesian feedback 
would be far more difficult to implement that Markovian feedback. 
Systematic linear approximations to Bayesian feedback fail even to match 
Markovian feedback for the two-level 
atom, so it is possible that inevitable experimental 
imperfections would unmake the general 
superiority of Bayesian feedback in this system. 
This is related to issues of robustness in classical 
control theory \cite{ZhoDoyGlo97}, which has only begun to be explored in 
in quantum systems \cite{Doh00,GamWis01}. 
Quite probably Markovian, Bayesian, and perhaps other forms of 
feedback will all have roles to play in the control of nonlinear quantum systems.

\section*{Acknowledgments}

HMW is grateful to Andrew Doherty and Kurt Jacobs for enlightening 
discussions. This work was supported by the Australian Research Council.

\end{multicols}

\begin{references}
\bibitem{fn2}
Note that when we describe the feedback as Markovian, we
mean that the resulting dynamics of the quantum system is Markovian.
This is a much stronger notion than that usually implied by the use of
the word Markovian in
classical control theory \cite{Jac93}, where it means that the dynamics
of the total system, including the monitoring device and feedback
loop, is Markovian.
\end{references}
\end{document}